\documentclass[prd,eqsecnum,preprint,showpacs,nofootinbib]{revtex4}
\usepackage{bm}
\usepackage{stmaryrd}
\usepackage{amsmath}
\usepackage{amssymb}
\usepackage{graphicx}
\usepackage{textcomp}
\usepackage{calrsfs}
\usepackage{yfonts}

\newcommand{\nn}{\nonumber}

\newcommand{\be}{\begin{equation}}
\newcommand{\ee}{\end{equation}}
\newcommand{\ben}{\begin{equation*}}
\newcommand{\een}{\end{equation*}}
\newcommand{\bea}{\begin{eqnarray}}
\newcommand{\eea}{\end{eqnarray}}

\newcommand{\bnabla}{\bm{\nabla}}
\newcommand{\bGamma}{\bm{\Gamma}}

\DeclareMathOperator{\Tr}{Tr}
\DeclareMathOperator{\tr}{tr}

\begin{document}

\title{Casimir-Polder repulsion near edges:
wedge apex and a screen with an aperture}

\date{\today}

\author{Kimball A. Milton}\email{milton@nhn.ou.edu}
\author{E. K. Abalo}\email{abalo@nhn.ou.edu}
\author{Prachi Parashar}\email{prachi@nhn.ou.edu}
\author{Nima Pourtolami}\email{nimap@ou.edu}
\affiliation{Homer L. Dodge Department of
Physics and Astronomy, University of Oklahoma, Norman, OK 73019-2061, USA}
\author{Iver Brevik}\email{iver.h.brevik@ntnu.no}
\author{Simen \AA. Ellingsen}\email{simen.a.ellingsen@ntnu.no}
\affiliation{Department of Energy and Process Engineering, Norwegian University
of Science and Technology, N-7491 Trondheim, Norway}

\begin{abstract}
Although repulsive effects have been predicted for quantum vacuum forces
between bodies with nontrivial electromagnetic properties, such as between
a perfect electric conductor and a perfect magnetic conductor, realistic 
repulsion seems difficult to achieve.  
Repulsion is possible if the medium between the bodies
has a permittivity in value intermediate to those of the two bodies, but this
may not be a useful configuration.  Here, inspired by recent numerical work,
we initiate analytic calculations of the 
Casimir-Polder interaction between an atom with
anisotropic polarizability and a plate with an aperture.  In particular,
for a semi-infinite plate, and, more generally, for a wedge,
the problem is exactly solvable, and for 
sufficiently large anisotropy, Casimir-Polder repulsion is indeed possible,
in agreement with the previous numerical studies. 
In order to achieve repulsion, what is needed is a sufficiently sharp
edge (not so very sharp, in fact) so that the directions of
polarizability of the conductor and the atom are roughly normal
to each other. The machinery for
carrying out the calculation with a finite aperture is presented.
As a motivation for the quantum 
calculation, we carry out the corresponding classical
analysis for the force between a dipole and a metallic sheet with a circular
aperture, when the dipole is on the symmetry axis and oriented in the same direction.
\end{abstract}

\maketitle
\section{Introduction}
There has been increasing interest in utilizing the quantum vacuum force or
the Casimir effect in nanotechnology employing mesoscopic objects
\cite{Rodriguez:2010zz}.  Although the original Casimir effect, between
parallel conducting or dielectric plates separated by vacuum 
\cite{casimir48, lifshitz56},  always gives an attractive force between the
plates, introducing a material (liquid) with an intermediate value of the 
dielectric constant can result in repulsion \cite{lifshitz61}, which
has now been observed \cite{capasso09}.  For precursors, see
\cite{milling96,meurk97,lee01,lee02,feiler08}.
[The first experimental test of the Lifshitz theory with an intermediate
liquid (helium) was that of Sabisky and Anderson \cite{sabisky};
application of the Lifshitz theory to the melting of water ice was
considered by Elbaum and Schick \cite{elbaum}.]
A recent experiment involving air bubbles in a liquid with boundary walls
is described in Ref.~\cite{tabor}.
However, this type of repulsion is unlikely
to have many applications in building devices.

There are well-known repulsive quantum forces in vacuum.  The first example
was found by Boyer \cite{boyer68}.  He computed the self-stress of a perfectly
conducting spherical shell due to quantum electrodynamic field fluctuations
and found a repulsive result, but the meaning of such a self-energy is
extremely obscure.  He later found \cite{boyer74}  a more observable effect, 
that the force between a perfect electrically conducting plane ($\varepsilon$, the
permittivity, goes to infinity) and a parallel perfect magnetic conducting plane
($\mu$, the permeability, goes to infinity) is repulsive.  This, again, may
be a difficult situation to approximately replicate in practice, because
the unusual magnetic properties must persist over a wide frequency range.

There has been extensive interest in designing metamaterials that could
give rise to Casimir repulsion by simulating a magnetic response
\cite{Henkel,Pirozhenko,Rosa,Rosa2,Zhao,comment}.
Despite some early optimism, the conclusion seems to have transpired that 
repulsion is impossible between metamaterials made from dielectric and 
metallic components \cite{Yannopapas, Silveirinha,McCauley}.  For
recent attempts using dielectric/magnetic setups see 
Refs.~\cite{maslovski,zeng,grushin}.  

Several years ago there was an interesting suggestion by Sopova and Ford
\cite{Sopova:2004qw} that the force between a small dielectric sphere
and a dielectric wall was oscillatory, so there were a number of repulsive
regimes.  However, this effect was canceled by plasmon modes leaving the
usual attractive result \cite{ford06}.  Earlier Ford had suggested \cite{ford}
that the frequency response of materials might be manipulated in order
to achieve repulsion, but this was proved to be impossible \cite{genet}.
%See also \cite{ellingsen08}.

Thus it was extremely interesting when Levin et al.\
showed examples of repulsion between conducting objects, in particular
between an elongated cylinder above a conducting plane with a circular
aperture \cite{Levin:2010zz}.  (An analytic counterpart is given in
\cite{maghrebi}.)
They first gave examples of repulsive
forces between arrays of electric dipoles, and an
electric dipole and a conducting plane with an aperture cut out.
Then they turned to quantum vacuum forces between conducting objects, computed
by quite impressive ``brute force'' finite-difference time-domain and
boundary-element methods.

The purpose of the present paper is to try to understand these phenomena
analytically.  We first show, in Sec.~\ref{sec2}, that there is no repulsion
possible in the weak coupling regime, where because the materials are dilute
one may sum Casimir-Polder interactions between atoms \cite{Casimir:1947hx}.
However, there is repulsion in classical electrostatics between a 
system of three dipoles (Sec.~\ref{sec3})  and between a fixed dipole and
a conducting plane with an aperture, which we discuss in Sec.~\ref{sec4},
both in two and three dimensions.
This is an interesting pedagogical problem, for it involves mixed coupled
integral equations, like those for an electrified disk, or a plane with an
aperture with different constant electric fields at large distances above
and below the punctured plane \cite{jackson}.  These problems exhibit closed
form solutions, and clearly exhibit repulsion when the dipole is directly
above the aperture and is sufficiently close.  In Sec.~\ref{sec5} we turn
to the real problem, that of the Casimir-Polder force between an anisotropic
polarizable atom and a punctured dielectric plane.  Because solving
the integral equations arising for the Green's dyadic 
for the plate with aperture is rather complicated,
in Sec.~\ref{sec6} we content ourselves with computing the Casimir-Polder
interaction between a polarizable atom and a perfectly conducting wedge.
When the opening angle of the wedge approaches $2\pi$, this describes the
interaction between an atom and a semi-infinite conducting plane.  We exhibit
situations in which repulsive forces in certain directions can arise for
anisotropic atoms, in qualitative agreement with numerical work
\cite{Levin:2010zz}.  In Appendix A we give another derivation of the Casimir-Polder
energy formula for the wedge, based on a closed form for the Green's dyadic,
and in Appendix B we give a classical calculation of a conducting ellipsoid
above a conducting plate with a circular aperture in the presence of a background
field.

A word about terminology: When we say ``atom'' we mean any microscopic particle which may 
be described by a polarisability tensor.  Our calculations assume that we are in the
retarded regime, so that static (frequency-independent) polarizabilities may be employed.
Should lower frequency transitions dominate (which could occur with some molecules), so
that the separations are in the non-retarded regime, electrostatic results are valid 
(but for a factor of 1/2---See Eq.~(\ref{energy}) below and Ref.~\cite{embook}).

In this paper we set $\hbar=c=1$, and all results are expressed in Gaussian
units except that Heaviside-Lorentz units are used for Green's dyadics.
\section{Weak Coupling Calculation}
\label{sec2}
\subsection{Scalar field}
\label{sec2a}
\begin{figure}
 \begin{center}
\includegraphics[scale=1]{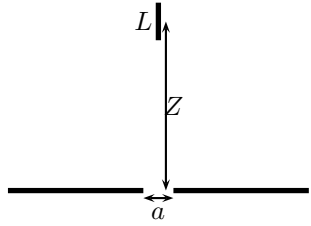}
\caption{\label{fig1} Two-dimensional geometry of a needle of
length $L$ a distance $Z$ above a line with a gap of width $a$.}
\end{center}
\end{figure}

We first illustrate the ideas by considering the case of a massless
 scalar field in two dimensions.  The quantum vacuum energy between two 
weakly coupled potentials
$V_1$ and $V_2$ is
\be
U_{12} =-\frac1{32\pi^2}\int (d\mathbf{r})(d\mathbf{r'})
\frac{V_1(\mathbf{r})V_2(\mathbf{r'})}{|\mathbf{r-r'}|^2},
\ee
the scalar analog of the Casimir-Polder force between atoms.
Here we consider the potentials as shown in Fig.~\ref{fig1},
which represents a needle of length $L$ on the symmetry axis
a distance $Z$ above a line with a gap of width $a$.
The potentials are given by
\begin{subequations}
\bea
 V_1(x,z)&=&\lambda_1 \delta (x)\theta(z-Z+L/2)\theta(Z+L/2-z),\\
V_2(x,z)&=&\lambda_2\delta(z)[\theta(x-a/2)+\theta(-x-a/2)].
\eea
\end{subequations}
This means that the interaction energy is
\bea
U_{12}=-\frac{\lambda_1\lambda_2}{32\pi^2}\int_{Z-L/2}^{Z+L/2}dz \left\{
\int_{a/2}^\infty+\int_{-\infty}^{-a/2}\right\} dx\frac1{x^2+z^2}.
\eea
To get the force on the needle, we simply have to integrate on $x$, and
differentiate with respect to the limits of the $z$ integral:
\be
 F=-\frac\partial{\partial Z}U_{12}=\frac{\lambda_1\lambda_2}{8\pi^2a}\left[
\frac{\arctan(2Z/a+L/a)}{2Z/a+L/a}-\frac{\arctan(2Z/a-L/a)}{2Z/a-L/a}
\right],
\ee
which, because $F<0$, always represents an attractive force between
the punctured line and the needle.  Note that although the force
vanishes at $Z=0$, the energy there, which represents the work done
in bringing the needle in from infinity, is not zero.

\subsection{Electromagnetic field}
\label{sec2b}
\begin{figure}
 \begin{center}
\includegraphics[scale=1]{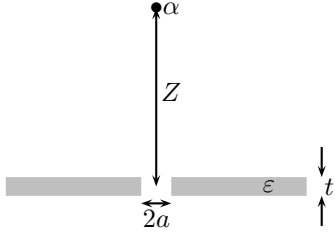}
\caption{\label{fig2} Three-dimensional geometry of a polarizable
atom a distance $Z$ above a dielectric slab with a circular
aperture of radius $a$.}
\end{center}
\end{figure}

Now we consider the quantum vacuum force between dilute dielectric
media, which may be obtained from the Casimir-Polder potential between
isotropic polarizable atoms \cite{Casimir:1947hx},
\be
U_{\rm CP}=-\frac{23}{4\pi}\alpha_1\alpha_2\frac1{r^7},\label{cp}
\ee
where $r$ is the distance between the atoms.
We might mention that equation (\ref{cp}) is in general valid in the 
retarded limit where the atomic polarizability can be regarded as constant.
(For more details, see the review \cite{buhmann-welsch}.)
The result is applicable provided that the 
atom-plate separation is much greater 
than the atomic transition wavelength (typically some hundreds of
nanometers for ground-state atoms). The media have
dielectric constants $\varepsilon_i=1+4\pi N_i\alpha_i$, where 
$N_i$ represents the density of atoms of type $i$.
Specifically, we consider a three-dimensional configuration, in which an
atom of isotropic polarizability $\alpha$ is placed on the
symmetry axis a distance $Z$ above a dielectric plate of thickness
$t$ with a circular hole in the middle, as shown in Fig.~\ref{fig2}.
  The quantum interaction energy is
\bea
U&=&-\frac{23}{(4\pi)^2}\alpha(\varepsilon-1)\int_{\rm slab}(d\mathbf{r})
\frac1{[(z-Z)^2+r_\perp^2]^{7/2}}\nn\\
&=&-\frac{23}{60\pi a^4}\alpha(\varepsilon-1)\left[
\frac{(t+2Z)(6a^2+(t+2Z)^2)}{(4a^2+(t+2Z)^2)^{3/2}}+(Z\to -Z)\right].
\eea
It is easy to see that the force $F=-\partial U/\partial Z$ is always negative,
i.e., attractive.

A more favorable case for possible repulsion would be an anisotropic
atom.  It is easy to derive the appropriate generalization of
the Casimir-Polder potential in this case, starting from the
weak-coupling multiple scattering formula \cite{brevikfest}
\be
U_{12}=\frac{i}2\Tr \bm{\Gamma}_0V_1\bm{\Gamma}_0 V_2,
\ee
where the free Green's dyadic is ($\zeta=-i\omega$)
\be
\bm{\Gamma}_0(\mathbf{r,r'})=(\bm{\nabla\nabla-1}\zeta^2)\frac{e^{-|\zeta||\mathbf{r-r'}|}}
{4\pi|\mathbf{r-r'}|}.\label{fgf}
\ee
Following the procedure given in Ref.~\cite{brevikfest}, 
we find for an isotropic medium facing an anisotropic atom
\be
U=\frac{\varepsilon-1}{32\pi^2}\int_{\rm slab}(d\mathbf{r})\frac1{|\mathbf{r-R}|^7}
\left[13 \tr \bm{\alpha}+7\frac{(\mathbf{r-R})\cdot\bm{\alpha}\cdot(\mathbf{r-R})}{(\mathbf{r-R})^2}
\right],
\ee
where $R=(0,0,Z)$ is the position of the atom, relative to the center of the aperture.
This may be easily checked to reduce to the usual Casimir-Polder result (\ref{cp}) when
$\bm{\alpha}=\alpha\bm{1}$.

Let's consider the extreme case when only $\alpha_{zz}$ is significant.  Then the integrals
may be easily carried out, with the result
\be
U=\frac{\alpha_{zz}(\varepsilon-1)}{60\pi a^4}\left[\frac{t+2Z}{[4+(t+2Z)^2]^{5/2}}
[156a^4+70a^2(t+2Z)^2+7(t+2Z)^4]+(Z\to-Z)\right].
\ee
This, again, always gives rise to an attractive force.  

An interesting special case is
when the aperture is small compared to the thickness of the dielectric.  Then
the energy is a step function,
\be
U=-\frac7{30\pi a^4}\alpha_{zz}(\varepsilon-1)\theta(t-2|Z|),\quad a\ll t,
\ee
which gives rise to a $\delta$-function force just when the atom enters and exits
the aperture.  If the aperture is very large compared to the thickness of the slab,
$t\ll a$, the energy and force are proportional to the thickness of the slab,
\be
U=-\frac{1}{80\pi a^4}\alpha_{zz}(\varepsilon-1)\frac{13a^2+18 Z^2}{(a^2+Z^2)^{7/2}}a^4
t.\ee

\section{Classical dipole interaction}
\label{sec3}
\begin{figure}
 \begin{center}
\includegraphics[scale=1]{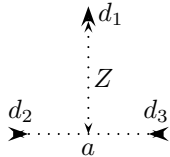}
\caption{\label{fig3} Configuration of three dipoles, two of which are
antiparallel, and one perpendicular to the other two.}
\end{center}
\end{figure}

It is possible to achieve a repulsive force between a configuration of
fixed dipoles.  Consider the situation illustrated in Fig.~\ref{fig3}.
Here we have two dipoles, of strength $d_2$ and $d_3$ lying along the $x$
axis, separated by a distance $a$.  A third dipole of strength $d_1$ lies
along the $z$ axis.  If the two parallel dipoles are oppositely directed
and of equal strength,
\be
\mathbf{d}_2=-\mathbf{d}_3=d_2 \mathbf{\hat x}, 
\ee
and are equally distant from the $z$ axis,
and the dipole on the $z$ axis is directed along that axis,
\be
\mathbf{d}_1=d_1\mathbf{\hat z},
\ee
the force on that dipole is along the $z$ axis:
\be
F_z=3ad_1d_2\frac{a^2/4-4 Z^2}{(Z^2+a^2/4)^{7/2}},
\ee
which changes sign at $Z=a/4$.  That is, for distances $Z$ larger than this,
the force is attractive (in the $-z$ direction) while for shorter distances
the force is repulsive (in the $+z$ direction).  Evidently, by symmetry,
the dipole-dipole energy vanishes at $z=0$.  Consistent with Earnshaw's
theorem, the point where the force vanishes is an unstable point with
respect to deviations in the $x$ direction.

In view of this self-evident finding, it might seem surprising that the 
interaction between a polarizable atom and a dilute medium (made up of 
polarizable atoms) studied in Sec.~\ref{sec2b} failed to exhibit a 
repulsive regime, but this is because the medium is isotropic.

\section{Classical interaction between a dipole and a conducting plane
with an aperture}\label{sec4}
In this section, we consider the interaction between a dipole and a perfectly
conducting plane containing an aperture.  We first consider two dimensions.
(As above, we denote the Cartesian coordinates by $x$ and $z$ for uniformity with
the three-dimensional situation.)
\subsection{Dipole above aperture in a conducting line}
\label{sec4a}

If we use the Green's function which vanishes on the entire line $z=0$,
\be
G(\mathbf{r,r'})=-\ln[(x-x')^2+(z-z')^2]+\ln[(x-x')^2+(z+z')^2],
\ee
so
\be G(x,0;x',z')=0,
\ee
we can calculate the electrostatic potential at any point above the $z=0$ plane
to be
\be
\phi(\mathbf{r})=\int_{z>0}(d\mathbf{r'})G(\mathbf{r,r'})\rho(\mathbf{r'})
+\frac1{4\pi}\int_{\rm ap}dS'\frac\partial{\partial z'}G(\mathbf{r,r'})\bigg|_{z'=0}
\phi(\mathbf{r'}),
\ee
where the volume integral is over the charge density of the dipole,
\be
\rho(\mathbf{r})=-\mathbf{d}\cdot\bm{\nabla}\delta(\mathbf{r-R}),\quad \mathbf{R}=(0,Z).
\ee
The surface integral extends only over the aperture because the potential
vanishes on the conducting sheet.  If we choose $\mathbf{d}$ to point along
the $z$ axis we easily find
\be
\phi(x,z>0)=2d\left[\frac{z-Z}{x^2+(z-Z)^2}+ \frac{z+Z}{x^2+(z+Z)^2}\right]
+\frac1\pi\int_{-a/2}^{a/2} dx'\frac{z}{(x-x')^2+z^2}\phi(x',0),
\label{potgf}
\ee where $a$ is the width of the aperture.

Now the free Green's function in two dimensions is
\be
G_0(\mathbf{r,r'})=4\pi\int\frac{(d\mathbf{k})}{(2\pi)^2}\frac{e^{ik_x(x-x')}
e^{ik_z(z-z')}}{k_x^2+k_z^2}
=\int_{-\infty}^\infty dk_x\frac1{|k_x|}e^{ik_x(x-x')}e^{-|k_x||z-z'|}.
\ee
Then the surface integral in Eq.~(\ref{potgf}) is
\be
\int_{-\infty}^\infty \frac{dk_x}{2\pi}e^{ik_x x}e^{-|k_x|z}\tilde\phi(k_x),
\ee
in terms of the Fourier transform of the field
\be
\tilde \phi(k_x)=\int_{-\infty}^\infty dx' e^{-ik_x x'}\phi(x',0)
=2\int_0^{a/2}dx'\cos k_xx'\phi(x',0),
\ee
since $\phi(x,0)$ must be an even function for the geometry considered.
Thus we conclude
\be
\phi(x,z>0)=2d\left[\frac{z-Z}{x^2+(z-Z)^2}+\frac{z+Z}{x^2+(z+Z)^2}\right]
+\frac1\pi\int_0^\infty dk\cos kx \,e^{-kz}\tilde\phi(k).
\ee
This becomes an identity as $z\to0$.

The electric field in the aperture is
\be
E_z(x,z=0+)=-\frac\partial{\partial z}\phi(x,z)\bigg|_{z-0+}=
-4d\frac{x^2-Z^2}{(x^2+Z^2)^2}+\frac1\pi\int_0^\infty dk\,k\cos kx\,\tilde\phi(k).
\ee

On the other side of the aperture, there is no charge density, so for
$z<0$ the potential is
\be
\phi(x,z<0)=\frac1\pi\int_0^\infty dk\cos kx \,e^{kz}\tilde\phi(k),
\ee
so the $z$-component of the electric field in the aperture is
\be
E_z(x,z=0-)=-\frac\partial{\partial z}\phi(x,z)\bigg|_{z=0-}
=-\frac1\pi\int_0^\infty dk\,k\cos kx\,\tilde\phi(k).
\ee
Because we require that the electric field be continuous in the aperture,
and the potential vanish on the conductor, we obtain the two coupled
integral equations for this problem,
\begin{subequations}
\bea
4d\frac{x^2-Z^2}{(x^2+Z^2)^2}&=&\frac2\pi\int_0^\infty dk\,k\cos kx\,
\tilde\phi(k),\quad 0<|x|<a/2,\\
0&=&\int_0^\infty dk\cos kx\,\tilde\phi(k),\quad |x|>a/2.
\eea
\end{subequations}

In fact, these equations have a simple solution \cite{Khanzhov}
\be
\tilde\phi(k)=-\frac{4Zd\pi}a\int_0^1dx\,x\frac{J_0(kax/2)}{(x^2+4Z^2/a^2)^{3/2}}.
\ee
From this, we can work out the energy of the system from
\be
U=-\frac12dE_z(0,Z)=\frac12d\frac{\partial\phi}{\partial z}\bigg|_{z=Z,x=0},
\label{energy}
\ee
where the factor of 1/2 comes from the fact that this must be the energy
required to assemble the system.  In computing this energy we must, of course,
drop the self-energy of the dipole due to its own field.  We are then left
with 
\bea 
U_{\rm int}&=&-\frac{d^2}{4Z^2}-\frac{d}{2\pi}\int_0^\infty dk\,k\, e^{-kZ}\tilde
\phi(k)\nn\\
&=&-\frac{d^2}{4Z^2}+Z^2d^2\left(\frac2a\right)^4\int_0^1\frac12 dx^2
\frac1{(x^2+4Z^2/a^2)^3}\nn\\
&=&-\frac{4Z^2d^2}{(a^2+4Z^2)^2},
\eea
where to get the second line we used the derivative of Eq.~(\ref{freegcc}).
This is exactly two times larger that the result quoted in 
Ref.~\cite{Levin:2010zz}.\footnote{This is not the factor of 1/2 in Eq.~(\ref{energy}).
It is not possible to trace the origin of the discrepancy, since the authors
of that reference merely quote the result.}
Since this vanishes at $Z=0$ and $Z=\infty$, the force must change from
attractive to repulsive, which happens at $Z=a/2$. 

\subsection{Three dimensional aperture interacting with dipole}\label{sec4b}
It is quite straightforward to repeat the above calculation in three
dimensions.  Again we are considering a dipole, polarized on the symmetry
axis, a distance $Z$ above a circular aperture of radius
$a$  in a conducting plate.  

The free three-dimensional
 Green's function in cylindrical coordinates has the representation
\be
\frac1{\sqrt{\rho^2+z^2}}=\int_0^\infty dk\,J_0(k\rho)e^{-k|z|},\label{freegcc}
\ee
and so if we follow the above procedure we find for the potential above
the plate
\bea
\phi(\mathbf{r_\perp},z>0)&=&d\left[\frac{z-Z}{[r_\perp^2+(z-Z)^2]^{3/2}}+
\frac{z+Z}{[r_\perp^2+(z+Z)^2]^{3/2}}\right]\nn\\
&&\quad\mbox{}+\int_0^\infty dk\,k\, e^{-kz}J_0(kr_\perp)\Phi(k),
\eea
where the Bessel transform of the potential in the aperture is
\be
\Phi(k)=\int_0^\infty d\rho\,\rho\, J_0(k\rho)\phi(\rho,0).
\ee
Thus the integral equations resulting from the continuity of the $z$-component
of the electric field in the aperture and the vanishing of the potential on
the conductor are
\begin{subequations}
\bea
d\frac{r_\perp^2-2Z^2}{[r_\perp^2+Z^2]^{5/2}}&=&\int_0^\infty dk \,k^2 J_0(k
r_\perp)\Phi(k),\quad r_\perp<a,\\
0&=&\int_0^\infty dk\,k J_0(kr_\perp)\Phi(k),\quad r_\perp>a.
\eea
\end{subequations}

The solution to these equations is given in Titchmarsh's book \cite{titchmarsh}, 
and after a bit of manipulation we obtain
\be
\Phi(k)=-\left(\frac{2ka}\pi\right)^{1/2}\frac{d}{ka}\int_0^1 dx\,x^{3/2}
J_{1/2}(xka)\frac{2Z/a}{(x^2+Z^2/a^2)^2}.
\ee
Then the energy (\ref{energy}) may be easily evaluated using
\be
\int_0^\infty dk\,k^{3/2}e^{-kZ}J_{1/2}(kax)=2\sqrt{\frac{2xa}\pi}
\frac{Z}{(x^2a^2+Z^2)^2}.
\ee
The energy can again be expressed in closed form:
\be
U=-\frac{d^2}{8Z^3}+\frac{d^2}{4\pi Z^3}\left[\arctan\frac{a}Z
+\frac{Z}a\frac{1+8/3(Z/a)^2-(Z/a)^4}{(1+Z^2/a^2)^3}\right].
\ee
This is always negative, but vanishes at infinity and at zero:
\be
Z\to 0:\quad U\to-\frac4{5\pi}d^2\frac{Z^2}{a^5}.
\ee
This means that for some value of $Z\sim a$ the force changes from
attractive to repulsive.  Numerically, we find that the force changes
sign at $Z=0.742358a$.

The reason why the energy vanishes when the dipole is centered in the
aperture is clear: Then the electric field lines are perpendicular to the
conducting sheet on the surface, and the sheet could be removed without
changing the field configuration.

Our goal is to analytically find the quantum (Casimir) analog of
this classical respulsion.

\section{Strong coupling---force between an atom and a punctured plane
dielectric}
\label{sec5}
Now we turn to the real problem.  Our starting point is the general
expression for the vacuum energy \cite{brevikfest}
\be
U=\frac{i}2\Tr\ln\bm{\Gamma\Gamma}_0^{-1},
\ee
where $\bm{\Gamma}$ is the full Green's dyadic for the problem, and 
$\bm{\Gamma}_0^{-1}$ is the inverse of the free Green's dyadic (\ref{fgf}), namely
\be
\bm{\Gamma}_0^{-1}=\frac1{\omega^2}\bm{\nabla}\times\bm{\nabla}\times-\bm{1}.
\ee
In the presence of a potential $\mathbf{V}$, the full Green's dyadic has
the symbolic form
\be
\bm{\Gamma}=(\bm{1}-\bm{\Gamma}_0\mathbf{V})^{-1}\bm{\Gamma}_0.
\label{Gammaexp}
\ee

Here we are thinking of the interaction between a dielectric medium,
characterized by an isotropic permittivity, so $V_1=\varepsilon-1$, and a polarizable
atom, represented by a polarizability dyadic, as shown in Fig.~\ref{fig2},
\be
\mathbf{V}_2=4\pi\bm{\alpha}\delta(\mathbf{r-R}),
\ee
where $\mathbf{R}$ is the position of the dipole.  We are only interested
in a single interaction with the latter potential, so we have for the interaction
energy
\be
U_{12}=\Tr \mathbf{V}_2\frac\delta{\delta V_1}\left[-\frac{i}2\ln\left(1-\bm{\Gamma}_0V_1
\right)\right]
=\frac{i}2\Tr\left(\bm{\Gamma}_1-\bm{\Gamma}_0\right)\mathbf{V}_2,
\ee
where we have used Eq.~(\ref{Gammaexp}) for the potential $V_1$ describing
 the dielectric slab plus aperture and we have
subtracted the term that represents the self-energy of the atom 
with its own field.
This subtraction happens automatically if we start from the ``$TGTG$'' form,
\be
U_{12}=-\frac{i}2\Tr\ln\left(\bm{1}
-\bGamma_1\mathbf{V}_1\bGamma_2\mathbf{V}_2\right)
\approx\frac{i}2\Tr \bGamma_1 \mathbf{V}_1\bGamma_0\mathbf{V}_2
=\frac{i}2\Tr \left(\bGamma_1-\bGamma_0\right)\mathbf{V}_2,
\ee
because $\mathbf{V}_2$ is weak.
This implies the Casimir-Polder expression for the interaction between 
the polarizable atom and the dielectric
\be
U_{\rm CP}=-\int_{-\infty}^\infty d\zeta \tr\bm{\alpha}\cdot(\bm{\Gamma}
-\bm{\Gamma}_0)(Z,Z).\label{alphagamma}
\ee

We could also derive this result from the formula for the force on a dielectric
body in an inhomogeneous electric field \cite{embook},
\be
\mathbf{F}=-\frac1{8\pi}\int (d\mathbf{r}) E^2(\mathbf{r})\bm{\nabla}\varepsilon,
\ee
which classically says that a dielectric body experiences a force pushing it into
the region of stronger field.  This implies the interaction energy
\be
U=-\frac12\alpha E^2(\bm{0},Z),
\ee and when we make the quantum-field-theoretic replacement,
\be
\frac1{4\pi}\langle \mathbf{E(r)E(r')}\rangle\to\frac1i\bm{\Gamma}(\mathbf{r,r'})
=\frac1{i}\int\frac{d\omega}{2\pi}\bGamma(\mathbf{r,r'};\omega),
\ee we recover the static isotropic version of Eq.~(\ref{alphagamma}) after the self-energy
is subtracted.

\subsection{No aperture}
When the aperture is not present, we are considering the well-studied case
of a dielectric slab, of thickness $t$, interacting with a polarizable atom.
Because the Green's dyadic in this situation, denoted $\bm{\Gamma}^{(0)}$, 
then possesses translational invariance in the $x$-$y$ plane,
we can express it in terms of a reduced Green's dyadic,
\be
\Gamma^{(0)}(\mathbf{r,r'})=\int\frac{(d\mathbf{k_\perp})}{(2\pi)^2}e^{i\mathbf{k_\perp
\cdot(r-r')_\perp}}\mathbf{g}(z,z';k_\perp).
\ee
In the case of an isotropic atom, the trace of the Green's dyadic occurs, which is
for the reduced Green's dyadic
\be
\tr \mathbf{g}(Z,Z)=-\zeta^2 g^H(Z,Z)+\left(\frac\partial{\partial Z}\frac\partial{\partial
Z'}+k_\perp^2\right)g^E(Z,Z')\bigg|_{Z'=Z},
\ee
in terms of the transverse electric (H) and transverse magnetic (E) Green's functions.
These subtracted quantities are for $z,z'$ above the dielectric
\be g^{H,E}(z,z')-g_0^{H,E}(z,z')=\frac1{2\kappa}R^{H,E}e^{-\kappa(z+z'-t)},
\ee in terms of the reflection coefficients
\begin{subequations}
\bea
R^H=\frac{\kappa-\kappa'}{\kappa+\kappa'}+4\frac{\kappa\kappa'}{\kappa^{\prime2}
-\kappa^2}\frac1D,\\
R^E=\frac{\kappa-\bar\kappa'}{\kappa+\bar\kappa'}+4\frac{\kappa\bar\kappa'}{
\bar\kappa^{\prime2}-\kappa^2}\frac1{\bar D},
\eea
\end{subequations}
where
\be
\kappa^2=k_\perp^2+\zeta^2,\quad \kappa^{\prime2}=k_\perp^2+\varepsilon\zeta^2,\quad
\bar\kappa^{\prime}=\kappa^{\prime}/\varepsilon
\ee
and 
\be
D=\left(\frac{\kappa+\kappa'}{\kappa-\kappa'}\right)^2e^{2\kappa' t}-1,
\ee
with $\bar D$ obtained from this by replacing $\kappa'$ by $\bar\kappa'$ except
in the exponent.  These results are rather trivially obtained by multiple
scattering arguments.

Now the interaction energy is
\bea
U&=&-\alpha \int_{-\infty}^\infty d\zeta\int\frac{(d\mathbf{k_\perp})}{(2\pi)^2}
\left[-\zeta^2 R^H+(2k^2+\zeta^2)R^E\right]\frac1{2\kappa}e^{-\kappa(2Z-t)}\nn\\
&=&-\frac\alpha{4\pi}\int_0^\infty d\zeta \int d k_\perp^2\frac1\kappa
e^{-2\kappa(Z-t/2)}\bigg\{(\varepsilon-1)\zeta^4\frac{e^{2\kappa't}-1}{(\kappa+\kappa')^2
e^{2\kappa't}-(\kappa-\kappa')^2}\nn\\
&&\quad\mbox{}+\frac{\varepsilon-1}{\varepsilon}(2k^2+\zeta^2)\left(k^2\left(1+\frac
1\varepsilon\right)+\zeta^2\right)\frac{e^{2\kappa't}-1}{(\kappa+\kappa'/\varepsilon)^2
e^{2\kappa't}-(\kappa-\kappa'/\varepsilon)^2}\bigg\}.
\eea
This is precisely the result found, for example, by Zhou and Spruch \cite{spruch}.

\subsection{Integral equations for Green's dyadic}
We now specialize to the case where the plane $z=0$ consists of a perfectly
conducting screen with a circular aperture of radius $a$ at the origin.
The Green's dyadic satisfies the differential equation
\be
\left(\frac1{\omega^2}\bnabla\times\bnabla\times-\bm{1}\right)\cdot\bGamma(\mathbf{
r-r'})=\bm{1}\delta(\mathbf{r-r'}),\label{gdif}
\ee
subject to the boundary conditions
\be
\mathbf{\hat z}\times \bGamma(\mathbf{r,r'})\bigg|_{|\mathbf{r}_\perp|>a,z=0}
=0,
\ee
which just states that the tangential components of the
 electric field must vanish on the conductor.
Following Levine and Schwinger \cite{levine-schwinger} we introduce auxiliary
electric and magnetic Green's dyadics $\bGamma^{(1,2)}(\mathbf{r,r'})$ which
satisfy the same differential equation (\ref{gdif}) but with the boundary
conditions satisfied on the entire $z=0$ plane:
\be
\mathbf{\hat z}\times\bGamma^{(1)}(\mathbf{r,r'})\bigg|_{z=0}=0,
\quad \mathbf{\hat z}\times(\bnabla\times\bGamma^{(2)}(\mathbf{r,r'}))
\bigg|_{z=0}=0.
\ee
These can be constructed in terms of the free Green's dyadic $\bGamma_0$, subject
only to outgoing boundary conditions at infinity, as given in Eq.~(\ref{fgf}),
\be
\bGamma_0(\mathbf{r,r'})=\left(\mathbf{1}\omega^2+\bnabla\bnabla
\right)G(|\mathbf{r-r'}|),
\ee
expressed in turn in terms of the Helmholtz Green's function
\be
G(R)=\frac{e^{i|\omega|R}}{4\pi R}.
\ee
We can write, after the Euclidean rotation $|\omega|\to i\zeta$, the free
Green's dyadic in the explicit form ($\mathbf{R}=\mathbf{r-r'}$)
\be
\bGamma_0(\mathbf{r,r'})=-\frac{G(R)}{R^2}\left[\bm{1}
\left(1+\zeta R+\zeta^2R^2\right)-\frac{\mathbf{R R}}{R^2}\left(3+3\zeta R+\zeta^2 
R^2\right)\right].
\ee

In terms of this last dyadic, the auxilliary Green's dyadics have the form
\be
z,z'>0:\quad \bGamma^{(1),(2)}(\mathbf{r,r'})=\bGamma^{(0)}(\mathbf{r,r'})
\mp\bGamma^{(0)}(\mathbf{r,r'}-2\mathbf{\hat z} z')
\cdot(\bm{1}-2\mathbf{\hat z\hat z}).
\ee

Now using Green's second identity, it is easy to prove
\begin{subequations}
\bea
\bnabla\times\bGamma^{(2)}(\mathbf{r,r'})&=&[\bnabla'\times \bGamma^{(1)}]^T(
\mathbf{r',r}),\\
\bGamma^{(1),(2)}(\mathbf{r,r'})&=&[\bGamma^{(1),(2)}]^T(\mathbf{r',r}),
\eea
\end{subequations}
where $T$ signifies transposition.  In the same way we may derive the integral
equations for the Green's dyadic for the screen with the aperture
\begin{subequations}
\bea
z,z'>0:\quad&& \bGamma(\mathbf{r,r'})=\bGamma^{(1)}(\mathbf{r,r'})
-\frac1{\zeta^2}
\int_{\rm ap} dS''\bnabla\times\bGamma_+^{(2)}(\mathbf{r,r''})\cdot
\mathbf{\hat z}\times \bGamma(\mathbf{r'',r'}),\nn\\
\\
z<0<z':\quad&& \bGamma(\mathbf{r,r'})=-
\frac1{\zeta^2}
\int_{\rm ap} dS''\bnabla\times\bGamma_-^{(2)}(\mathbf{r,r''})\cdot
\mathbf{\hat z}\times \bGamma(\mathbf{r'',r'}),
\eea
\end{subequations}
where the $\pm$ subscripts on $\bGamma^{(2)}$ 
indicate that the Green's function
is defined in the domain above or below the $z=0$ plane.  The continuity of the
$z$-component of the electric field in the aperture then leads to the integral
equation
\be
\mathbf{\hat z}\cdot \bGamma^{(1)}(\mathbf{r,r'})\bigg|_{z\to 0+}
=\frac1{\zeta^2}
\int_{\rm ap} dS'' \mathbf{\hat z}\cdot \bnabla \times(\bGamma_+^{(2)}
+\bGamma_-^{(2)})(\mathbf{r,r''})\cdot\mathbf{\hat z}\times \bGamma(
\mathbf{r'',r'}).
\ee

The system of integral equations defining the Green's dyadic is rather more
complicated that those describing the corresponding (classical) static
potential problem, so we will defer the discussion of strategies for its
solution to a subsequent publication.  We will here turn to a situation
that can be solved exactly.

\section{Casimir-Polder Force between Atom and a Conducting Wedge}
\label{sec6}
\begin{figure}
 \begin{centering}
\includegraphics[scale=1]{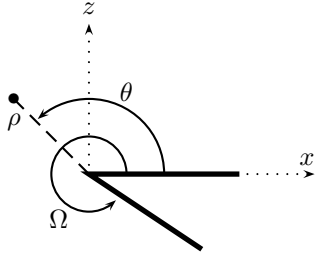}
\caption{\label{figwedge}  Polarizable atom, located at polar coordinates
$\rho$, $\theta$, within a conducting wedge with dihedral angle $\Omega$.}
 \end{centering}

\end{figure}

The interaction between a polarizable atom and a perfectly conducting
half-plane is a special case of the vacuum interaction between such an atom
and a conducting wedge.  For the case of an isotropic atom, this was 
considered by Brevik, Lygren, and Marachevsky \cite{blm}.  (This followed
on earlier work by Brevik and Lygren \cite{bl} and DeRaad and Milton 
\cite{dm}.)
In terms of the opening dihedral angle of the wedge $\Omega$, 
which we describe
in terms of the variable $p=\pi/\Omega$, the electromagnetic Green's
dyadic has the form (here the translational direction is denoted by $y$, 
and one plane of the wedge lies in the $z=0$ plane, the other intersecting 
the $xz$ plane on the line $\theta=\Omega$---see Fig.~\ref{figwedge})
\bea
\bGamma(\mathbf{r,r'})&=&2p\sum_{m=0}^\infty{}'\int\frac{dk}{2\pi}
\bigg[-\bm{\mathcal{M}}\bm{\mathcal{M}}^{\prime*}(\nabla_\perp^2-k^2)\frac1{\omega^2}
F_{mp}(\rho,\rho')\frac{\cos mp\theta \cos mp\theta'}\pi e^{ik(x-x')}  \nn\\
&&\quad \mbox{}+\bm{\mathcal{N}}\bm{\mathcal{N}}^{\prime*}\frac1{\omega}
G_{mp}(\rho,\rho')\frac{\sin mp\theta \sin mp\theta'}\pi e^{ik(x-x')}\bigg].
\eea
The first term here refers to TE (H) modes, the second to TM (E) modes.
The prime on the summation sign means that the $m=0$ term is counted with
half weight.  In the polar coordinates in the $xz$ plane, $\rho$ and $\theta$,
the H and E mode operators are
\begin{subequations}\label{hemodeops}
\bea
\bm{\mathcal{M}}&=&\bm{\hat \rho}\frac\partial{\rho\partial\theta}
-\bm{\hat\theta}\frac\partial{\partial \rho},\\
\bm{\mathcal{N}}&=&ik\left(\bm{\hat \rho}\frac\partial{\partial\rho}
+\bm{\hat\theta}\frac\partial{\rho\partial \theta}\right)
-\mathbf{\hat y}\nabla_\perp^2,
\eea
\end{subequations}
where the transverse Laplacian is
\bea
\nabla_\perp^2=
\frac1\rho\frac\partial{\partial \rho}\rho\frac\partial{\partial\rho}
+\frac1{\rho^2}
\frac{\partial^2}{\partial \theta^2}.
\eea
In this situation, the boundaries are entirely in planes of constant $\theta$,
so the radial Green's functions are equal to the free Green's function
\be
\frac1{\omega^2}F_{mp}(\rho,\rho')=\frac1\omega G_{mp}(\rho,\rho')
=-\frac{i\pi}{2\lambda^2}
J_{mp}(\lambda \rho_<)H^{(1)}_{mp}(\lambda \rho_>),
\ee
with $\lambda^2=\omega^2-k^2$.
We will immediately make the Euclidean rotation, $\omega\to i\zeta$, where
$\lambda\to i\kappa$, $\kappa^2=\zeta^2+k^2$, so the free Green's functions 
become $-\kappa^{-2}I_{mp}(\kappa\rho_<)K_{mp}(\kappa \rho_>)$.

We start by considering the most favorable case for CP repulsion, where the
atom is only polarizable in the $z$ direction, that is, only $\alpha_{zz}\ne0$.
In the static limit, then the only component of the Green's dyadic that
contributes is 
\bea
\int\frac{d\zeta}{2\pi}\Gamma_{zz}
&=&\frac{2p}{4\pi^3}\int dk\,d\zeta
\bigg\{\left[\zeta^2\sin^2\theta\sin^2mp\theta-k^2\cos^2\theta\cos^2mp\theta
\right]\frac{m^2p^2}{\kappa^2\rho_<\rho_>}I_{mp}(\kappa\rho_<)K_{mp}(\kappa
\rho_>)\nn\\
&&\quad\mbox{}-\left[k^2\sin^2\theta\sin^2 mp\theta-\zeta^2\cos^2\theta
\cos^2mp\theta\right]I'_{mp}(\kappa\rho_<)K'_{mp}(\kappa \rho_>)\bigg\}.
\eea
Here we note that the off diagonal $\rho$-$\theta$ terms in $\bGamma$ cancel.
We have regulated the result by point-splitting in the radial coordinate.
At the end of the calculation, the limit $\rho_<\to\rho_>=\rho$ is to be
taken.

Now the integral over the Bessel functions is given by
\be
\int_0^\infty d\kappa\,\kappa\, I_\nu(\kappa\rho_<)K_\nu(\kappa\rho_>)
=\frac{z^\nu}{\rho_>^2(1-\xi^2)},
\ee
where $\xi=\rho_</\rho_>$. After that the $m$ sum is easily carried out
by summing a geometrical series.  Care must also be taken with the $m=0$ term
in the cosine series.  The result of a straightforward calculation leads to
\be
\label{vacuum}
\int\frac{d\zeta}{2\pi}\Gamma_{zz}=-\frac{\cos 2\theta}{\pi^2\rho^4}
\frac1{(\xi-1)^4}+\mbox{finite},
\ee
where the divergent term, as $\xi\to1$,
 may, through a similar calculation, be shown to
be that corresponding to the vacuum in absence of the wedge, that is, that
obtained from the free Green's dyadic.  Therefore, we must subtract this term
off, leaving for the static Casimir energy (\ref{alphagamma})
\be
U^{zz}_{\rm CP}=-\frac{\alpha_{zz}(0)}{8\pi}\frac1{\rho^4\sin^4p\theta}
\left[p^4-\frac{2}3p^2(p^2-1)\sin^2p\theta+\frac{(p^2-1)(p^2+11)}{45}\sin^4p\theta
\cos2\theta\right].\label{ucpzz}
\ee
This result is derived by another method in Appendix \ref{app1}.

A small check of this result is that as $\theta\to 0$ (or $\theta\to\Omega$)
we recover the expected Casimir-Polder result for an atom above an infinite
plane:
\be
U_{\rm CP}^{zz}\to -\frac{\alpha_{zz}(0)}{8\pi Z^4},
\ee
in terms of the distance of the atom above the plane, $Z=\rho\theta$.
This limit is also obtained when $p\to1$, for when $\Omega=\pi$ we are
describing a perfectly conducting infinite plane.

A very similar calculation gives the result for
an isotropic atom, $\bm{\alpha}=\alpha\bm{1}$, which was first given in
Ref.~\cite{blm}:
\be
U_{\rm CP}=-\frac{3\alpha(0)}{8\pi\rho^4\sin^4p\theta}\left[p^4-\frac23p^2(p^2-1)
\sin^2p\theta-\frac13\frac1{45}(p^2-1)(p^2+11)\sin^4 p\theta\right].
\ee
Note that this is not three times $U_{\rm CP}^{zz}$ in Eq.~(\ref{ucpzz})
 because the $\cos 2\theta$ factor in the last term in the latter is replaced
by $-1/3$ here.  This case was reconsidered recently, for example, in
Ref.~\cite{mendez}.
\begin{figure}
 \begin{centering}
\includegraphics[scale=1]{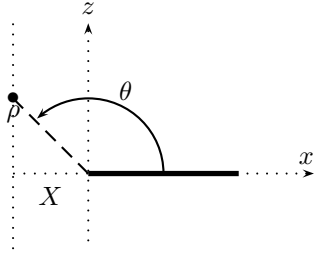}
\caption{\label{fighp}  Polarizable atom, above a half conducting plane,
free to move on a line perpendicular to the plane but a distance $X$ to
the left of the plane.}
 \end{centering}
\end{figure}

\subsection{Repulsion by a conducting half-plane}
Let us consider the special case $p=1/2$, that is $\Omega=2\pi$, the case of
a semi-infinite conducting plane. This was the situation considered,
for anisotropic atoms, in recent papers  by Eberlein and Zietal
\cite{ce1,ce2,ce3}.  Note that in such a case, for the completely
anisotropic atom, $U_{\rm CP}^{zz}=0$ at $\theta=\pi/2$, that is, there is no
force on the dipole when it is polarized perpendicular to the half-sheet
and directly above the edge, as observed in Refs.~\cite{ce2,ce3}.
 
 Consider a particle free to move along a
line parallel to the $z$ axis, a distance $X$ to the left of the 
semi-infinite plane.  See Fig.~\ref{fighp}.  The half-plane $x<0$
constitutes an aperture of infinite width. With $X$ fixed, we
can describe the trajectory by $u=X/\rho=-\cos\theta$, 
which variable ranges from zero to one.  The polar angle is given by 
\be
\sin^2\frac\theta2=\frac{1+u}2.
\ee
 The energy for an isotropic atom is given by
\be
U_{\rm CP}=-\frac{\alpha(0)}{32\pi}\frac1{X^4}V(u),
\ee
where 
\be
V(u)=3u^4\left[\frac1{(1+u)^2}+\frac1{u+1}+\frac14\right].\label{uiso}
\ee
The energy for the completely anisotropic atom is
\be
V_{zz}=\frac13V(u)+\frac{u^4}2(1-3u^2).
\ee
If we consider instead a cylindrically symmetric polarizable
atom in which
\be
\bm{\alpha}=\alpha_{zz}\mathbf{\hat z\hat z}+\gamma\alpha_{zz}(\mathbf{\hat x
\hat x+\hat y\hat y})=\alpha_{zz}(1-\gamma)\mathbf{\hat z\hat z}
+\gamma\alpha_{zz}\bm{1},\label{gammapol}
\ee
where $\gamma$ is the ratio of the transverse polarizability to the
longitudinal polarizability of the atom. 
Then the effective potential is
\be
(1-\gamma)V_{zz}+\gamma V,\label{effpot}
\ee
and the $z$-component of the force on the atom is
\be
F^\gamma_z=-\frac{\alpha_{zz}(0)}{32\pi}\frac1{X^5}u^2\sqrt{1-u^2}\frac{d}{du}
\left[\frac12u^4(1-\gamma)(1-3u^2)+\frac13(1+2\gamma)V(u)\right],\label{fz}
\ee
where $V$ is given by Eq.~(\ref{uiso}).  Note that the energy (\ref{effpot}), or the
quantity in square brackets in Eq.~(\ref{fz}), only vanishes at $u=1$ (the
plane of the conductor) when $\gamma=0$. Thus, the argument given in 
Ref.~\cite{Levin:2010zz} applies only for the completely anisotropic case.

\begin{figure}
 \begin{center}
\includegraphics[scale=1]{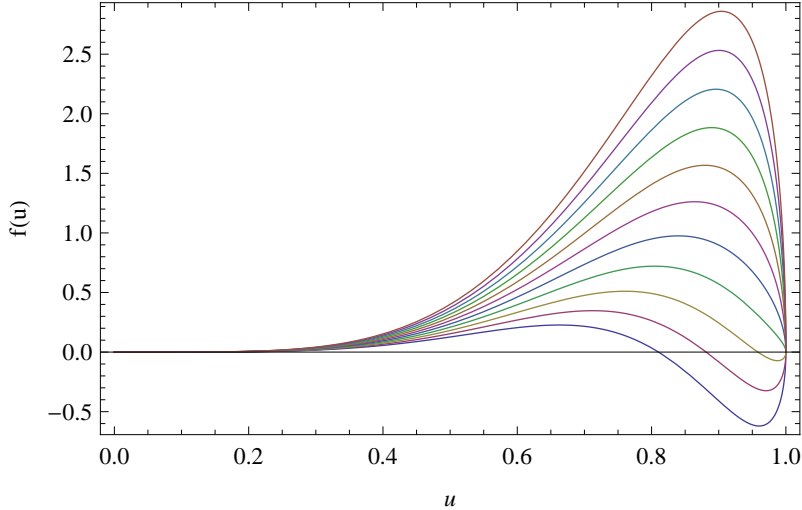}
\caption{\label{repsemi} (Color online) 
The $z$-component of the force between an anisotropic
atom (with ratio of transverse to longitudinal polarizabilities $\gamma$)
and a semi-infinite perfectly conducting plane, $z=0$, $x>0$.
$F_z=-\alpha_{zz}/(32\pi X^5) f(u)$ in terms of the variable 
$u=X/\rho=-\cos\theta$. Here
the atom lies on the line $x=0$, $y=-X$, and $\rho$ is the distance from the
edge of the plane and the atom.  Here, $f>0$ corresponds to an attractive
force on the $z$ direction, and $f<0$ corresponds to a repulsive force.
The different curves correspond to different values of $\gamma$, $\gamma=0$ to
1 by steps of 0.1, from bottom to top.
For $\gamma<1/4$ a repulsive regime always occurs when the atom is sufficiently
close to the plane of the conductor.} 
\end{center}
\end{figure}

\begin{figure}
\begin{center}
\includegraphics[scale=1]{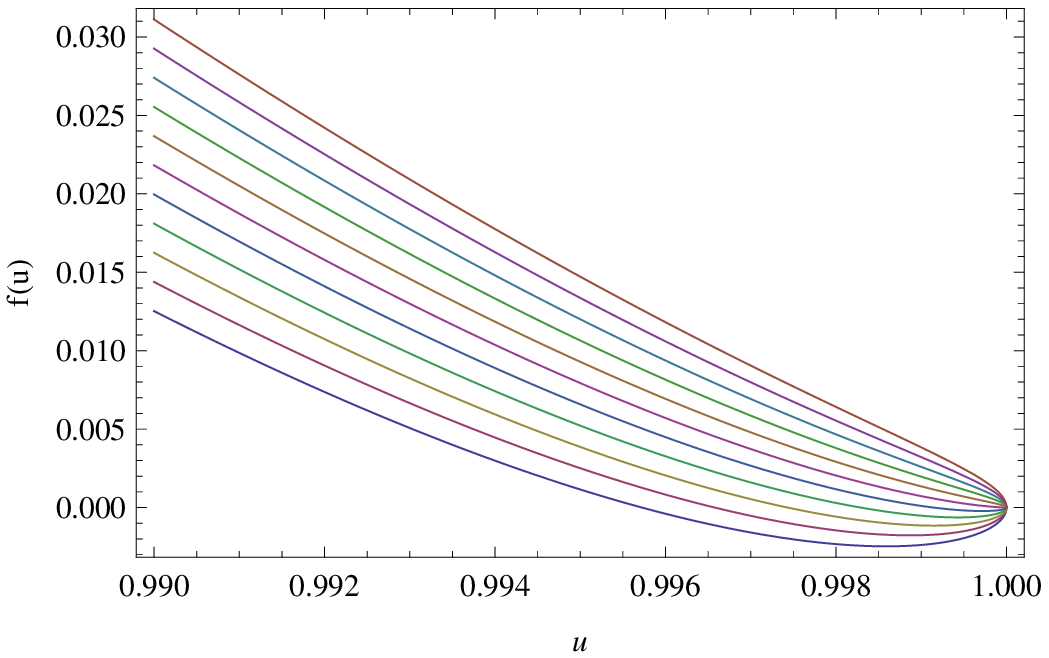}
\caption{\label{repsemi2} (Color online)  Same as Fig.~\ref{repsemi}.  The region close
to the plane, $1\ge u\ge 0.99$,
 with $\gamma$ near the critical value of 1/4.
Here from bottom to top are shown the results for values of $\gamma$ from
0.245 to 0.255 by steps of 0.001.}
\end{center}
\end{figure}

The force is plotted in Figs.~\ref{repsemi}, \ref{repsemi2}.  
It will be seen that if $\gamma$
is sufficiently small, when the atom is sufficiently close to
the plane of  the plate the
$z$-component of the force is repulsive rather than attractive.  The critical
value of $\gamma$ is $\gamma_c=1/4$.  This is a completely analytic exact
analog of the numerical calculations shown in Ref.~\cite{Levin:2010zz},
where the interaction was considered
between a conducting plane with an aperture (circular
hole or slit), and a conducting cylindrical or ellipsoidal object.
Our calculation demonstrates that three-body effects are not required to
exhibit Casimir-Polder repulsion.

It is interesting to observe that the same critical value of $\gamma$
occurs for the nonretarded regime of a circular aperture, as follows from
a simple computation based on the result of Ref.~\cite{ce3}.
For example, applying the result there for an atom with polarizability 
given by Eq.~(\ref{gammapol}) placed a distance $Z$ along the symmetry axis of an circular
aperture of radius $a$ in a conducting plane gives an energy
\bea
U&=&-\frac1{16\pi^2}\int_{-\infty}^\infty d\zeta\,\alpha_{zz}(\zeta)\nn\\
&&\times\frac1{Z^3}\left\{(1+\gamma)\left(\frac\pi2+\arctan\frac{Z^2-a^2}{2aZ}\right)
+\frac{2aZ}{(Z^2+a^2)^3}\left[(1+\gamma)(Z^4-a^4)-\frac83(1-\gamma)a^2Z^2\right]\right\}.\nn\\
\eea
It is easy to see that this has a minimum for $z>0$, and hence there is a repulsive
force close to the aperture, provided $\gamma<\gamma_c=1/4$.

\subsection{Repulsion by a wedge}
\begin{figure}
\begin{center}
\includegraphics[scale=1]{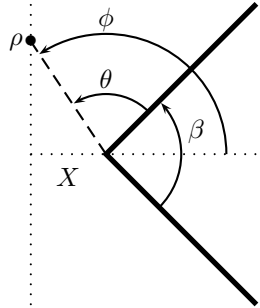}
\caption{\label{cpwedge}  A polarizable atom outside a perfectly
conducting wedge of interior angle $\beta$.  
The atom is located at polar angles $\rho$,
$\phi$ relative to the symmetry plane of the wedge.}
\end{center}
\end{figure}

It is very easy to generalize the above result for a wedge, $p>1/2$.  That is,
we want to consider a strongly anisotropic atom, with only $\alpha_{zz}$
significant, to the left of a wedge of opening angle
\be
\beta=2\pi-\Omega,
\ee
as shown in Fig.~\ref{cpwedge}.  
We want the $z$ axis to be perpendicular to the symmetry axis of the wedge
so the relation between the polar angle of the atom and the
angle to the symmetry line
is 
\be
\phi=\theta+\beta/2,
\ee
where, as before, $\theta$ is the angle relative to the top surface of the
wedge.  Then, it is obvious that the formula for the Casimir-Polder energy
(\ref{ucpzz}) is changed only by the replacement of $\cos2\theta$ by 
$\cos2\phi$, with no change in $\sin p\theta$.
Now we can ask how the region of repulsion depends on the wedge angle $\beta$.

\begin{figure}
\begin{center}
\includegraphics[scale=1]{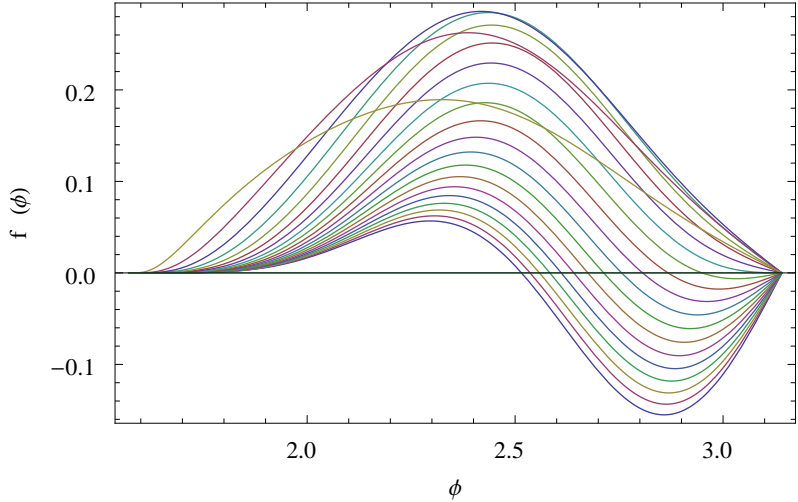}
\caption{\label{fig:cpwedge} (Color online) The $z$-component of the force on an
completely anisotropic atom moving on a line perpendicular to a
wedge. The different curves are for various values of $\beta$ from
0 to $\pi$ by steps of $\pi/20$, from bottom up.  The last few values
of $\beta$ have a markedly different character from the others.}
\end{center}
\end{figure}

Write for an atom on the line $x=-X$
\be
U^{zz}_{\rm CP}=-\frac{\alpha_{zz}(0)}{8\pi X^4}V(\phi),
\ee
where 
\be
V(\phi)=\cos^4\phi\left[\frac{p^4}{\sin^4\frac\pi2\frac{\phi-\beta/2}{\pi
-\beta/2}}-\frac23\frac{p^2(p^2-1)}{\sin^2\frac\pi2\frac{\phi-\beta/2}{\pi
-\beta/2}}+\frac1{45}(p^2-1)(p^2+11)\cos2\phi\right].
\ee
At the point of closest approach,
\be
V(\pi)=\frac1{45}(4p^2-1)(4p^2+11),
\ee
so the potential vanishes at that point only for the half-plane case,
$p=1/2$.  The force in the $z$ direction is
\begin{subequations}
\bea
F_z&=&-\frac{\alpha_{zz}}{8\pi}\frac1{X^5}f(\phi),\\
f(\phi)&=&\cos^2\phi\frac{\partial V(\phi)}{\partial\phi}.
\eea
\end{subequations}
Fig.~\ref{fig:cpwedge} shows the force as a function of $\phi$ for fixed $X$.
It will be seen that the force has a repulsive region for angles close
enough to the apex of the wedge, provided that the wedge angle is not too
large.  The critical wedge angle is actually
rather large, $\beta_c=1.87795$, or about
108$^\circ$.  For larger angles, the $z$-component of the force exhibits only
attraction.  Of course, the force is zero for $\beta=\pi$ because then
the geometry is translationally invariant in the $z$ direction.

\section{Conclusions}
This paper may be thought of as a counterpart to
Ref.~\cite{Levin:2010zz}.  While that reference proceeded on the basis of
numerical calculations, we have used analytic
approaches.  After some examples indicating that Casimir-Polder attraction
is typical, and always seems to occur in weak coupling,  we 
demonstrate that the quantum-vacuum Casimir-Polder interaction for
a sufficiently anisotropic atom above a conducting half plane can exhibit
regimes of repulsive forces for motion confined to certain specified 
directions.  This directly translates into repulsion between such an atom
and a plane with an aperture for motion along 
a line perpendicular to the plane. More complete analysis of that case
will be presented elsewhere. 

As we were putting finishing touches on this paper, Ref.~\cite{ce3} appeared,
which demonstrates in the {\it nonretarded\/} (van der Waals) regime,
repulsion could occur between an anisotropically polarizable atom and a
conducting plate with an aperture.  The critical value of the anisotropy is
the same as found here.

Perhaps most remarkable here is that not only can we achieve repulsion
with a half-plane, but also with a wedge geometry, even when the interior
angle of the wedge is greater than 90$^\circ$.  This indicates that while
anisotropy in both the atom and the conductor must be present for repulsion,
the anisotropy in the latter need not be too extreme, and that repulsion
in other geometries may be readily achievable.  Three-body forces are not
required, nor is a high degree of symmetry, as was present in 
Refs.~\cite{Levin:2010zz,ce3}.
\acknowledgments
We thank the US Department of Energy, and the US National Science Foundation,
for partial support of this research.  The support of the ESF Casimir Network
is also acknowledged. S.\AA.E. thanks Stefan Buhman, Stefan Scheel, and 
Alexander McCauley for discussions.  P.P acknowledges the hospitality of the
University of Zaragoza.

\appendix

\section{Derivation of anisotropic wedge CP force from closed-form Green's function}
\label{app1}
Many years ago Lukosz gave a closed form for the Green's functions for a 
perfectly conducting wedge \cite{lukosz}.  The four-dimensional Euclidean
 Green's dyadic has the closed form
\be
\bm{\Gamma}(\tau-\tau',x-x', \rho,\rho',\phi,\phi')
=-\mathbf{MM'}G^{\rm H}+\mathbf{NN'}G^{\rm E},
\ee
where the transverse differential operators are [cf.~Eq.~(\ref{hemodeops})]
\be
\mathbf{M}=\bm{\hat\rho}\frac1\rho\frac\partial{\partial\phi}
-\bm{\hat\phi}\frac\partial{\partial \rho}\equiv \bm{\mathcal{M}},
\quad 
\mathbf{N}=\bm{\hat\rho}\frac\partial{\partial\rho}
+\bm{\hat\phi}\frac1\rho\frac\partial{\partial\phi},
\ee
where there is an additional contribution to $\mathbf{N}$ in the $x$ direction.
This Green's dyadic is the frequency Fourier transform of that discussed in
Sec.~\ref{sec6}. Here the E (TM) and H (TE) Green's functions have the form
\be
G^{\rm H,E}=\chi(x,\rho,\tau;x',\rho',\tau';\phi-\phi')\pm
\chi(x,\rho,\tau; x'\rho',\tau';\phi+\phi'-\beta),
\ee
for a wedge of dihedral angle $\Omega$, with $\phi\in [-\Omega/2,\Omega/2]$.
Here
\be
\chi(x,\rho,\tau;x',\rho',\tau';\psi)=
\frac1{8\pi\Omega\rho\rho'\sinh\upsilon}\frac{\sinh(\pi\upsilon/\Omega)}{
\cosh(\pi\upsilon/\Omega)-\cos(\pi\psi/\Omega)},
\ee
where 
\be
\sinh\frac\upsilon2=\frac12\left[\frac{(\tau-\tau')^2+(x-x')^2+(\rho-\rho')^2}{\rho\rho'}
\right]^{1/2}.
\ee

For the interaction with an atom possessing only an $\alpha_{zz}$ 
polarizability, we need
\bea\label{gammazzcf}
\Gamma_{zz}&=&\cos(\phi+\phi')\left(\frac1{\rho\rho'}
\frac\partial{\partial\phi}
\frac\partial{\partial\phi'}-\frac\partial{\partial\rho}
\frac\partial{\partial\rho}
\right)\chi(\phi-\phi')\nn\\
&&\quad\mbox{}+2\left(\sin\phi\cos\phi'\frac1\rho\frac\partial{\partial\phi}
\frac\partial{\partial\rho'}
+\sin\phi'\cos\phi\frac1{\rho'}\frac\partial{\partial\phi'}
\frac\partial{\partial\rho}\right)\chi(\phi-\phi')\nn\\
&&\quad\mbox{}-\cos(\phi-\phi')\left(\frac1{\rho\rho'}
\frac\partial{\partial\phi}
\frac\partial{\partial\phi'}+\frac\partial{\partial\rho}
\frac\partial{\partial\rho'}\right)
\chi(\phi+\phi'-\Omega).
\eea
Here, we have suppressed all the arguments in $\chi$ except for the angular 
ones.
For our application here, we are interested in the coincidence limit, 
so from the outset we can set $\tau=\tau'$ and $x=x'$.  Then
\be
\sinh\frac\upsilon2=\frac12\frac{1-\xi}{\sqrt{\xi}},\quad \xi=\frac{\rho_<}{\rho_>},
\ee
which implies
\be
\upsilon=-\ln \xi.
\ee
Now we expand first in $\phi-\phi'$, 
then after the differentiations set $\phi=\phi'$,
and then expand in $\upsilon$, that is, in $1-\xi$. We immediately note that the
mixed derivative term in Eq.~(\ref{gammazzcf}) 
does not contribute, because there is no linear term in $\phi-\phi'$.
 The result of a straightforward
calculation is
\bea\label{gzzex}
\Gamma_{zz}&=&-\frac{\cos2\theta}{16\pi^2\rho^4}\left\{\frac{16}{(1-\xi)^4}
-\frac1{45}(p^2-1)(p^2+11)\right\}\nn\\
&&\quad\mbox{}+\frac1{16\pi^2\rho^4}\left\{\frac{p^4}{\sin^4 p\theta}
-\frac23\frac{p^2(p^2-1)}{\sin^2p\theta}\right\},
\eea
where $p=\pi/\Omega$, and we have switched to the angle from the ``upper'' 
plate, $\theta=\phi+\Omega/2$, which is chosen to run from $0$ to $\Omega$.
 The first term  in Eq.~(\ref{gzzex}) corresponds to the 
$\chi(\phi-\phi')$ contribution, and the second to the 
$\cos(\phi+\phi'-\Omega)$ contribution.
Note, the divergent term (as $\xi\to1$) 
is precisely the vacuum term given in Eq.~(\ref{vacuum}), and should be
subtracted off, and the rest, when multiplied by $-2\pi\alpha_{zz}$,
coincides with Eq.~(\ref{ucpzz}).

\section{ Electrostatic aspects: Conducting ellipsoid outside a conducting plate 
with a circular hole}

Consider a conducting uncharged solid ellipsoid with semiaxes $c>a>b$, centered at 
$X=Y=Z=0$. The ellipsoid is orientated such that the major semiaxis $c$ lies 
along the $Z$ axis. To describe the electrostatic potential $\phi$ in the 
external region, one can make use of ellipsoidal coordinates $\xi, \eta, \zeta$, 
corresponding to solutions for $u$ of the cubic equation
\begin{equation}
\frac{Z^2}{c^2+u}+\frac{X^2}{a^2+u}+\frac{Y^2}{b^2+u}=1. \label{1}
\end{equation}
The coordinate intervals are
\begin{equation}
\infty >\xi \geq -b^2, \quad -b^2 \geq \eta \geq -a^2,\quad -a^2\geq \zeta 
\geq -c^2.
\label{2}
\end{equation}
The relationships between the ellipsoidal and the Cartesian coordinates are 
given in
Ref.~\cite{landau-lifshitz}
%Landau-Lifshitz Vol. VIII, section 4
and will not be reproduced here.  We shall however need the line element,
\begin{equation}
dl^2=h_1^2\,d\xi^2+h_2^2\,d\eta^2+h_3^2\,d\zeta^2, \label{3}
\end{equation}
where
\begin{equation}
h_1=\frac{1}{2R_\xi} \sqrt{(\xi-\eta)(\xi-\zeta)}, \quad h_2=\frac{1}{2R_\eta}
\sqrt{(\eta-\zeta)(\eta-\xi)}, \label{4}
\end{equation}
\begin{equation}
h_3=\frac{1}{2R_\zeta}\sqrt{(\zeta-\xi)(\zeta-\eta)}, \quad R_u^2
=(u+c^2)(u+a^2)(u+b^2), \label{5}
\end{equation}
with $u=\xi,\eta,\zeta$.

In the following we assume axial symmetry around the $Z$ axis. Then 
$a\rightarrow b$, $\eta \rightarrow -b^2 $, and the equation for the 
surface of the ellipsoid becomes
\begin{equation}
\frac{Z^2}{c^2}+\frac{R^2}{b^2}=1, \label{6}
\end{equation}
with $R^2=X^2+Y^2$. We now have
\begin{equation}
Z=\pm \left[\frac{(\xi+c^2)(\zeta+c^2)}{c^2-b^2}\right]^{1/2}, \quad R=\left[ 
\frac{(\xi+b^2)(\zeta+b^2)}{b^2-c^2}\right]^{1/2}. \label{7}
\end{equation}
The ellipsoidal coordinates $\xi, \eta, \zeta$ reduce in the case of axisymmetry to 
so-called  prolate spheroidal coordinates $\xi$ and $\zeta$, lying in the intervals
\begin{equation}
\infty > \xi \geq -b^2,
\quad -b^2 \geq \zeta \geq -c^2. \label{8}
\end{equation}
Surfaces of constant $\xi$ and $\zeta$ are prolate spheroids and hyperboloids of 
revolution, the surfaces intersecting orthogonally. On the $Z$ axis ($R=0$) 
one has $\zeta =-b^2, Z=\pm \sqrt{\xi+c^2}$, whereas in the $XY$ plane ($Z=0$) one 
has $\zeta=-c^2, R=\sqrt{\xi+b^2}$. On the surface of the ellipsoid, $\xi=0$.

In free space outside the ellipsoid the Laplace equation reads
\begin{equation}
\nabla^2 \phi \equiv \frac{4}{\zeta-\xi}\left[\frac{R_\xi}{\xi+b^2}\,\frac{\partial}
{\partial \xi}\left(R_\xi \frac {\partial \phi}{\partial \xi}\right) -\frac{R_\zeta}
{\zeta+b^2}\,\frac{\partial}{\partial \zeta}\left(R_\zeta \frac{\partial \phi}
{\partial \zeta}\right) \right]=0. \label{9}
\end{equation}
Assume now that the ellipsoid is placed in an external potential $\phi_0$, 
axisymmetric with respect to the $Z$ axis so that $\phi_0=\phi_0(\xi,\zeta)$. We 
write the resulting potential $\phi$ in the form
\begin{equation}
\phi(\xi,\zeta)=\phi_0(\xi,\zeta)[1+F(\xi)], \label{10}
\end{equation}
so that $\phi_0 F$ is the perturbation of the external field. As the boundary 
condition $\xi=0$ on the surface has to hold for all values of $\zeta$, it is 
natural to make the ansatz that $F$ depends on $\xi$ only.

Inserting Eq.~(\ref{10}) into Eq.~(\ref{9})  we find that the terms containing $F$ 
as a factor sum up to zero, the reason being the validity of Eq.~(\ref{9}) also when 
$\phi$ is replaced by $\phi_0$. The remaining  terms containing $F'(\xi)$ and 
$F''(\xi)$ yield the equation
\begin{equation}
\frac{d^2 F}{d\xi^2}+\frac{dF}{d\xi}\frac{d}{d\xi}\ln\left( R_\xi \,\phi_0^2\right)=
0. \label{11}
\end{equation}
When integrating this equation, in order to preserve the validity of the ansatz 
$F=F(\xi)$, the coordinate $\zeta$ in $\phi_0$ has to be regarded as a parameter. 
The integration thus has to extend from $\xi=0$ (the surface) in the outward 
direction, along a line on the hyperboloid $\zeta$ = constant.

The solution of Eq.~(\ref{11}) can be written as
\begin{equation}
F=A\int_\xi^\infty \frac{d\xi}{R_\xi \phi_0^2}\,, \label{12}
\end{equation}
where the constant $A$ is determined from the condition $F(0)=-1$ on the ellipsoid 
surface. That means,
\begin{equation}
\phi=\phi_0\left[ 1-\frac{\int_\xi^\infty \frac{d\xi}{R_\xi\phi_0^2}}{\int_0^\infty 
\frac{d\xi}{R_\xi\phi_0^2}} \right]. \label{13}
\end{equation}
We now specify the form of $\phi_0$, as the potential from a grounded conducting 
plate lying in the $xy$ plane, when far  from the plate there are constant 
electric fields, directed normal to the plate, having different values on either 
side. In the plate there is a circular opening with radius $a$ (this radius not to 
be confused with the  semiaxis $a$ mentioned above). The center of the opening is 
at position $x=y=z=0.$ It is known (Ref.~\cite{jackson}, Sec.~3.13)
%(cf. Jackson 1999, sect. 3.13)
that on the $z$ axis
\begin{equation}
\phi_0(z)=\Phi_{00}\left[ 1-\frac{|z|}{a}\arctan \frac{a}{|z|} \right], \label{14}
\end{equation}
where $\Phi_{00}$ is a constant. At the origin, $\phi_0=\Phi_{00}$. At infinity, 
$|z|\rightarrow \infty$, $\phi_0\rightarrow 0$.

The center of the vertically oriented ellipsoid is at position $z=z_0$. Thus 
$z=z_0+Z$. We will assume that the ellipsoid is so slender that the variation of 
$\phi_0$ in the transverse $x$ and $y$ directions can be neglected. Thus we adopt 
the expression (\ref{14}) in the external field region of interest, 
$\phi_0=\phi_0(\xi,\zeta)$, $\xi$ and $\zeta$ being restricted to the same intervals
(\ref{8}) as before.

We consider now the upper half of the ellipsoid, $z\geq z_0$ or $Z\geq 0$. The 
nonperturbed potential, called $\phi_{0+}$,  is then
\begin{equation}
\phi_{0+} =\Phi_{00}\left[1-\frac{z_0+\sqrt{\xi+c^2}}{a}\arctan \frac{a}{z_0+
\sqrt{\xi+c^2}}\right]. \label{15}
\end{equation}
Thus the potential $\phi_+$ in Eq.~(\ref{13}) can  be found numerically, 
inserting $\phi_{0+}$ together with $R_\xi=(\xi+b^2)\sqrt{\xi+c^2}$. [In practice  
the following expansion can here be useful
\cite{NIST}%(Abramowitz-Stegun, formula 4.4.49):
\begin{equation}
\frac{1}{x}\arctan x=1+\sum_{k=1}^8a_{2k}x^{2k}+ {\cal{O}}(10^{-8}), \quad 0\leq x
\leq 1, \label{16}
\end{equation}
with coefficients $a_{2k}$ of order unity or less.]

The induced surface charge density $\sigma_+$ on the ellipsoid is
\begin{equation}
\sigma_+= -\left[\frac{\epsilon_0}{h_1}\frac{\partial \phi_+}
{\partial \xi}\right]_{\xi=0}
=-\left[\frac{2\epsilon_0 bc}{\sqrt{-\zeta}}\,\frac{\partial \phi_+}{\partial \xi}
\right]_{\xi=0}, \label{17}
\end{equation}
since on the surface $h_1=(b/2R_\xi)\sqrt{-\zeta}=(1/2bc)\sqrt{-\zeta}$. In view of 
the relationships between the ellipsoidal and Cartesian coordinates this can be 
reexpressed as
\begin{equation}
\sigma_+=-2\epsilon_0 \left[ \frac{Z^2}{c^4}+\frac{R^2}{b^4}\right]^{-1/2}\left[ 
\frac{\partial \phi_+}{\partial \xi}\right]_{\xi=0}. \label{18}
\end{equation}
From Eq.~(\ref{13}) it follows that the derivative $[\partial \phi_{0+}/\partial 
\xi]_{\xi=0}$ does not contribute to $\sigma_+$ (recall that $F(0)=-1$). The 
remaining term is
\begin{equation}
\left[ \frac{\partial \phi_+}{\partial \xi}\right]_{\xi=0}=\frac{1}{b^2c}\frac{1}
{[\phi_{0+}]_{\xi=0}}\left[\int_0^\infty \frac{d\xi}{R_\xi \,\phi_{0+}^2}
\right]^{-1}. \label{19}
\end{equation}
Thus for $z \geq z_0$ we get as solution
\begin{equation}
\sigma_+= \frac{\sigma_{0+}}{c} \left[ \frac{Z^2}{c^4}+\frac{R^2}{b^4}\right]^{-1/2},
\label{20}
\end{equation}
where $\sigma_{0+}$ is the constant
\begin{equation}
\sigma_{0+}=
-\frac{2\epsilon_0}{b^2}\,\frac{1}{\Phi_{00}}\,\frac{\left[
 \int_0^\infty \frac{d\xi}{R_\xi \,\phi_{0+}^2}\right]^{-1}}
{\left[1-\frac{z_0+c}{a}\arctan \frac{a}{z_0+c}\right]}\,
 \label{21}
\end{equation}
(recall again that $a$ is the radius of the hole). The dependence of $\sigma_+$ on 
the coordinates $Z$ and $R$ in Eq.~(\ref{20})  is actually the same  as for a charged 
ellipsoid in free space \cite{landau-lifshitz}.
The surface force density on the ellipsoid is $(\sigma^2/2\epsilon_0)\bf n$, $\bf n$
being the outward normal. The slope of the tangent to the surface is $dZ/dR=
-(c^2/b^2)R/Z$;  the slope of $\bf n$ is accordingly $(b^2/c^2)Z/R$. Denoting this as 
$\tan \theta$, we get when going over to ellipsoidal coordinates,
\begin{equation}
\tan \theta=\frac{b}{c}\left[ \frac{\zeta+c^2}{-\zeta-b^2}\right]^{1/2}. \label{22}
\end{equation}
The component of $\bf n$ along the $Z$ axis is then
\begin{equation}
n_Z= \sin
\theta=\frac{b}{\sqrt{c^2-b^2}}\left[\frac{\zeta+c^2}{-\zeta}\right]^{1/2},
\label{23}
\end{equation}
and we can now find the total vertical force $F_{Z+}$ on the upper
half of the ellipsoid by integrating over the actual surface. The
 line element along the meridian is
\begin{equation}
h_3d\zeta=\frac{1}{2}\left[\frac{\zeta}{(\zeta+b^2)(\zeta+c^2)}\right]^{1/2}
d\zeta, \label{24}
\end{equation}
and the surface element $dA$ becomes
\begin{equation}
dA=2\pi R \,h_3\,d\zeta=\frac{\pi
b}{\sqrt{c^2-b^2}}\left[\frac{-\zeta}{\zeta+c^2}\right]^{1/2}
d\zeta. \label{25}
\end{equation}
As $\sigma_+$ in Eq.~(\ref{20}) can be reexpressed as
\begin{equation}
\sigma_+=\sigma_{0+} \,\frac{b}{\sqrt{-\zeta}}, \label{26}
\end{equation}
we can calculate $F_{Z+}$ as
\begin{equation}
F_{Z+}=\int_{Z\geq 0}\frac{\sigma_+^2}{2\epsilon_0}n_Z
\, dA=\frac{\sigma_{0+}^2}{2\epsilon_0}\frac{\pi
b^4}{c^2-b^2}\int_{b^2}^{c^2}
\frac{d(-\zeta)}{(-\zeta)}=\frac{\sigma_{0+}^2}{\epsilon_0}\frac{\pi
b^4}{c^2-b^2}\ln \frac{c}{b}. \label{27}
\end{equation}
The expression is positive as expected; the force is acting upwards. The only 
dependence on the position $z_0$ lies in $\sigma_{0+}$, as $\sigma_{0+}=\sigma_{0+}(z_0)$ 
according to Eq.~(\ref{21}).

The lower half of the ellipsoid, $Z<0$, can be treated in an analogous way. 
A complicating element is here the presence of the conducting plate in the $xy$ plane, 
for radii $\rho \geq  a$. It means that we can no longer  extend the integration 
over $\xi$ in the solution (\ref{12})  to infinity in a straightforward way. 
We observe that the  undisturbed potential in the $xy$ plane can be written as
\begin{equation}
\phi_0(\rho,0)=
\left\{ \begin{array}{ll}
\Phi_{00}\sqrt{1-\rho^2/a^2},  & \rho \leq a \\
0,                               & \rho >a,
\end{array}
\right. \label{28}
\end{equation}
where $\rho^2=x^2+y^2$, $\Phi_{00}$ being the potential at the center.

Our approach will be based on the following two assumptions:

1) The $\xi$ integration will be terminated on the $xy$ plane, this implying that the
effect of the perturbation is assumed to be small at that level. This approximation 
is expected to be good except when the distance between the lower end of the 
ellipsoid and the plane is small.

2) Secondly, the integration over $\xi$ will be assumed to run over trajectories 
lying close to the $z$ axis, corresponding to $\zeta=-b^2$. This assumption simplifies 
the mathematical analysis. It is  supported by physical considerations also, since 
when the ellipsoid is slender the hyperboloids $\zeta=$ constant emerging from the 
surface of the ellipsoid near its lower end become concentrated in the vicinity of the $z$ 
axis.

As according to Eq.~(\ref{7}) the plane position $z=0$ in general corresponds to
\begin{equation}
z_0=\left[ \frac{(\xi+c^2)(\zeta+c^2)}{c^2-b^2}\right]^{1/2}, \label{29}
\end{equation}
our approximations imply that the $\xi$ integration is terminated at
\begin{equation}
\xi_{\rm plane}=z_0^2-c^2, \label{30}
\end{equation}
i.e., the same constant for the whole lower half of the ellipsoid.

As solution for the perturbed potential  we thus get
\begin{equation}
\phi_-=\phi_{0-}\left[ 1-\frac{\int_\xi^{\xi_{\rm plane}} \frac{d\xi}{R_\xi \,
\phi_{0-}^2}}{\int_0^{\xi_{\rm plane}} \frac{d\xi}{R_\xi \,\phi_{0-}^2}} \right], \label{31}
\end{equation}
where
\begin{equation}
\phi_{0-} =\Phi_{00}\left[1-\frac{z_0-\sqrt{\xi+c^2}}{a}\arctan \frac{a}{z_0
-\sqrt{\xi+c^2}}\right]. \label{32}
\end{equation}
The force $F_{Z-}$ on the lower half can now be calculated. As before, $R_\xi
=(\xi+b^2)\sqrt{\xi+c^2}$. Equation (\ref{26}) becomes replaced by
\begin{equation}
\sigma_-=\sigma_{0-} \frac{b}{\sqrt{-\zeta}}, \label{33}
\end{equation}
 where now
 \begin{equation}
 \sigma_{0-}=-\frac{2\epsilon_0}{b^2}\frac{1}{\Phi_{00}}\frac{ \left[
\int_0^{\xi_{\rm plane}}\frac{d\xi}{R_\xi \, \phi_{0-}^2}\right]^{-1}}{ 
\left[ 1-\frac{z_0-c}{a}\arctan \frac{a}{z_0-c} \right] }. \label{34}
 \end{equation}
The total force on the ellipsoid becomes
 \begin{equation}
 F_Z=F_{Z+}+F_{Z-}=\frac{\sigma_{0+}^2-\sigma_{0-}^2}{\epsilon_0}
 \frac{\pi b^4}{c^2-b^2}\ln \frac{c}{b}, \label{35}
 \end{equation}
 which can be rewritten as
\bea 
 F_Z &=& \frac{4\pi \epsilon_0}{\Phi_{00}^2}\frac{1}{c^2-b^2} \Bigg\{ 
\frac{\left[
 \int_0^\infty \frac{d\xi}{R_\xi \,\phi_{0+}^2}\right]^{-2}}
{\left[1-\frac{z_0+c}{a}\arctan \frac{a}{z_0+c}\right]^2}   \nn\\
&&\quad\mbox{}-\frac{ \left[\int_0^{\xi_{\rm plane}} \frac{d\xi}{R_\xi\, 
\phi_{0-}^2} \right]^{-2}}
{\left[ 1-\frac{z_0-c}{a}\arctan \frac{a}{z_0-c}\right]^2} \Bigg\} \ln 
\frac{c}{b}. \label{36}
\eea
In the limiting case of a sphere, $b \rightarrow c$, the expression becomes somewhat 
simpler,
\begin{equation}
F_Z=\frac{2\pi \epsilon_0}{\Phi_{00}^2}\frac{1}{c^2}\left\{ \frac{ \left[\int_0^\infty 
\frac{d\xi}{(\xi+c^2)^{3/2}\,\phi_{0+}^2}\right]^{-2}}{\left[1-\frac{z_0+c}{a}
\arctan \frac{a}{z_0+c}\right]^2}
-  \frac{ \left[\int_0^{\xi_{\rm plane}} \frac{d\xi}{(\xi+c^2)^{3/2}\,\phi_{0-}^2}
\right]^{-2}}{\left[1-\frac{z_0-c}{a}\arctan \frac{a}{z_0-c}\right]^2} \right\}. 
\label{37}
\end{equation}

We have made some numerical checks of these expressions (using Maple). They indicate 
that there is no change in the sign of the
force for various input parameters for the geometry. The force is attractive, as 
expected. It turns out that the dependence on the upper integration limit 
$\xi_{\rm plane} = z_0^2-c^2 $  is weak, as anticipated above.


\begin{thebibliography}{99}


\bibitem{Rodriguez:2010zz}
  A.~W.~Rodriguez, A.~P.~McCauley, D.~Woolf {\it et al.},
%``Nontouching Nanoparticle Diclusters Bound by Repulsive and Attractive Casimir Forces,''
  Phys.\ Rev.\ Lett.\  {\bf 104}, 160402 (2010).
  
\bibitem{casimir48} H. B. G. Casimir, Proc. Kon. Ned. Wetensch. {\bf 51},
793 (1948).

\bibitem{lifshitz56} E. M. Lifshitz, Zh. Eksp. Teor. Fiz. {\bf 29}, 94 (1956)
[Engl.~transl.: Sov. Phys. JETP {\bf 2}, 73 (1956)].

\bibitem{lifshitz61} I. D. Dzyaloshinskii, E. M. Lifshitz, and L. P. 
Pitaevskii, Usp. Fiz. Nauk {\bf 73}, 381 (1961) [Engl.~transl.: Sov. Phys. Usp.
{\bf 4}, 153 (1961)].

\bibitem{capasso09} J. Munday, F. Capasso, and V. A. Persegian, Nature 
{\bf 457}, 170 (2009).

\bibitem{milling96} A.~Milling, P.~Mulvaney, and I.~Larson, J.\ Colloid Interf.\ Sci.\ {\bf 180}, 460 (1996).
\bibitem{meurk97} A.~Meurk, P.~F.\ Luckham, and L.~Bergstrom, Langmuir {\bf 13}, 3896 (1997).
\bibitem{lee01} S.\ Lee and W.~M.\ Sigmund, J.\ Colloid Interf.\ Sci.\ {\bf 243}, 365 (2001).
\bibitem{lee02} S.\ Lee and W.~M.\ Sigmund, J.\ Colloids Surf.\ A {\bf 204}, 43 (2002).
\bibitem{feiler08} A.~A.\ Feiler, L.\ Bergstr\"{o}m, and M.~W.\ Rutland, Langmuir {\bf 24}, 2274 (2008).

\bibitem{sabisky} E. S. Sabisky and C. H. Anderson,
Phys.\ Rev.\ A {\bf7}, 790--806 (1973).

\bibitem{elbaum} M. Elbaum and M. Schick,
Phys.\ Rev.\ Lett.\ {\bf66}, 1713--1716 (1991).
\bibitem{tabor}
R. F. Tabor, R. Manica, D. Y. C. Chan, F. Grieser, and R. R. Dagastine, 
Phys.\ Rev.\ Lett.\  {\bf106}, 064501 (2011).

 


\bibitem{boyer68} T. H. Boyer, Phys. Rev. {\bf 174}, 1764 (1968).

\bibitem{boyer74} T. H. Boyer, Phys. Rev. A {\bf 9}, 2078 (1974).



\bibitem{Henkel} C. Henkel and K. Joulain, Europhys.\ Lett.\ {\bf72}, 929 (2005).

\bibitem{Pirozhenko} I. G. Pirozhenko and A. Lambrecht, J. Phys.\ A {\bf41}, 164015 (2008).

\bibitem{Rosa} F. S. S. Rosa, D. A. R. Dalvit, and P. W. Milonni, 
Phys.\ Rev.\ Lett.\ {\bf100}, 183602 (2008).

\bibitem{Rosa2} F. S. S. Rosa, D. A. R. Dalvit, and P. W. Milonni,
Phys.\ Rev.\ A {\bf78}, 032117 (2008).

\bibitem{Zhao} R. Zhao, J. Zhou, Th. Koschny, E. N. Economou, and C. M. Soukoulis, 
 Phys.\ Rev.\ Lett.\ {\bf103}, 103602 (2009).

\bibitem{comment} M. G. Silveirinha and S. I. Maslovski,
Phys.\ Rev.\ Lett.\ {\bf105}, 189301 (2010);  
R. Zhao, J. Zhou, Th. Koschny, E. N. Economou, and C. M. Soukoulis,
Phys.\ Rev.\ Lett.\ {\bf105}, 189302 (2010).

%These authors are careful in their optimism to start with (except perhaps for Zhau et al), and with good reason, it %seems, for it appears also this kind of repulsion is impossible, at least for a passive (thermal equilibrium) system:

\bibitem{Yannopapas} V. Yannopapas and N. V. Vitanov, Phys.\ Rev. Lett.\ {\bf103}, 120401 (2009).

\bibitem{Silveirinha} M. G. Silveirinha and S. I. Maslovski, Phys.\ Rev.\ A {\bf82}, 052508 (2010).

\bibitem{McCauley} A. P. McCauley et al., Phys.\ Rev.\ B {\bf82}, 165108 (2010).

\bibitem{maslovski}
S. I. Maslovski and M. G. Silveirinha, Phys.\ Rev.\ A {\bf83}, 022508 (2011).

\bibitem{zeng}
R. Zeng and Y. Yang, Phys.\ Rev.\ A {\bf83}, 012517 (2011).

\bibitem{grushin}
A. G. Grushin and A. Cortijo, Phys.\ Rev.\ Lett.\ {\bf106}, 020403 (2011).

 






\bibitem{Sopova:2004qw}
  V.~Sopova and L.~H.~Ford,
  %``Casimir force between a small dielectric sphere and a dielectric wall,''
  Phys.\ Rev.\  {\bf A70}, 062109 (2004).

\bibitem{ford06} L. H. Ford, J. Phys.\ A {\bf 39}, 6365 (2006).
 
\bibitem{ford} L. H. Ford, Phys.\ Rev.\ A {\bf48}, 2962--2967 (1993).

\bibitem{genet} C. Genet, A. Lambrecht, and S. Reynaud, Phys.\ Rev.\ A {\bf67}, 043811 (2003).

%\bibitem{ellingsen08} 
%Frequency spectrum of the Casimir force: Interpretation and a paradox
%S. \AA. Ellingsen, Europhys.\ Lett.\ {\bf82}, 53001 (2008).

\bibitem{Levin:2010zz}
  M.~Levin, A.~P.~McCauley, A.~W.~Rodriguez, M. T. Homer Reid, S. G. Johnson, 
  %``Casimir Repulsion between Metallic Objects in Vacuum,''
  Phys.\ Rev.\ Lett.\  {\bf 105}, 090403 (2010).
  
\bibitem{maghrebi}
%Diagrammatic expansion of the Casimir energy in multiple reflections: 
%Theory and applications
M. F. Maghrebi, Phys.\ Rev.\ D {\bf83}, 045004 (2011). 

\bibitem{Casimir:1947hx}
  H.~B.~G.~Casimir and D.~Polder,
  %``The Influence of retardation on the London-van der Waals forces,''
  Phys.\ Rev.\  {\bf 73}, 360-372 (1948).



\bibitem{jackson} J.D. Jackson, {\it Classical Electrodynamics} (Wiley, New York,
1999).

\bibitem{embook} J. Schwinger, L. L. DeRaad, Jr., K. A. Milton, and
W.-y. Tsai, {\it Classical Electrodynamics} (Perseus/Westview, New York,
1998).

\bibitem{buhmann-welsch} S.Y. Buhmann and D.-G. Welsch, Prog.\ Quantum Electron.\
{\bf 31}, 51 (2007).

\bibitem{brevikfest}
K. A. Milton,
P. Parashar and J. Wagner, in {\it The Casimir Effect and Cosmology}, 
ed. S. D. Odintsov, E. Elizalde, and O. B. Gorbunova, 
in honor of Iver Brevik (Tomsk State Pedagogical University) pp. 107-116 
(2009) [arXiv:0811.0128].

\bibitem{Khanzhov} A. D. Khanzhov, Inzherero-Fizicheskia Zhurnal {\bf 11}, 658
(1966) [Engl.~Transl.: J. Eng. Phys. Thermophys. {\bf 11}, 370 (1966)].

\bibitem{titchmarsh}
E. C. Titchmarsh, {\it Theory of Fourier Integrals} (Oxford, 1948).



\bibitem{spruch} F. Zhou and L. Spruch, Phys. Rev. A {\bf 52}, 297 (1995).

\bibitem{levine-schwinger} H. Levine and J. Schwinger, Comm.\ Pure Appl.\
Math.\ III 4, 355 (1950), reprinted in K. A. Milton and J. Schwinger,
{\it Electromagnetic Radiation: Variational Methods, Waveguides, and
Accelerators} (Springer, Berlin, 2006), p. 543.

%\bibitem{johnson} M. Levin, A. P. McCauley, A. W. Rodriguez, M. T. Homer Reid,
%and S. G. Johnson, Phys. Rev. Lett. {\bf 105}, 090403 (2010).
\bibitem{blm} I. Brevik, M. Lygren, and V, N. Marachevsky, Ann.\ Phys.\
(N.Y.) {\bf 267}, 134 (1998).

\bibitem{bl} I. Brevik and M. Lygren, Ann.\ Phys.\ (N.Y.) {\bf 251}, 157 
(1996).

\bibitem{dm} L. L. DeRaad, Jr., and K. A. Milton, 
Ann.\ Phys.\ (N.Y.) {\bf 136}, 229 (1981).

\bibitem{mendez} T. N. C. Mendez, F. S. S. Rosa, A. Ten\'orio, and C. Farina,
J. Phys.\ A: Math.\ Theor.\ {\bf 41}, 164020 (2008).

\bibitem{ce1} C. Eberlein and R. Zietal, Phys.\ Rev.\ A {\bf 75}, 032516 
(2007).

\bibitem{ce2} C. Eberlein and R. Zietal, Phys.\ Rev.\ A {\bf 80}, 012504
(2009).

\bibitem{ce3} C. Eberlein and R. Zietal, arXiv:1103.2381.

\bibitem{lukosz} W. Lukosz, Z. Phys.\ {\bf 262}, 327 (1971).

\bibitem{landau-lifshitz} L. D. Landau and E. M. Lifshitz,
{\it Electrodynamics of Continuous Media} (Pergamon, Oxford, 1960),
Sec.~4, p.~20ff.

\bibitem{NIST} M. Abramowitz and I. A. Stegun (Eds.) 
{\it Handbook of Mathematical Functions with Formulas, Graphs, and
Mathematical Tables},
National Bureau of Standards Applied Mathematics Series
(U.S. Government Printing Office, Washington, D.C., 1964), formula 4.4.49.
    
\end{thebibliography}
\end{document}